\pdfoutput=1
\documentclass[12pt]{iopart}

\usepackage{iopams}

\usepackage{epstopdf}
\usepackage{psfrag}
\usepackage{ps4pdf}

\usepackage{multirow}
\usepackage{paralist}
\usepackage{color}

\begin{document}

\newcommand{\D}{\mathrm{d}}
\newcommand{\pr}{{}^{\tiny{\mbox{��}}}}
\newcommand{\nuint}{\nu_{\mbox{\footnotesize{���}}}}
\newcommand{\Vreg}{V_{\mbox{\footnotesize{���}}}}

\newcommand{\OB}[2]
{\overbrace{\hspace{#1mm}}^{ #2}}

\newcommand{\zd}
{
\scriptstyle{b-z(\tau)}\;
 \Bigg\{
}

\newcommand{\BC}{
\OB{22}{\frac{b}{c}}
}

\newcommand{\ZC}{
\OB{10}{\frac{b-z(\tau)}{c}}
}

\title{Can corner-cube absolute gravimeters sense the effects of Special Relativity? 
}
\author{V D Nagornyi}
\address{Metromatix, Inc., 111B Baden Pl, Staten Island, NY 10306, USA}
\ead{vn2@member.ams.org}
\author{Y M Zanimonskiy}
\address{Institute of Radio Astronomy, National Academy of Sciences of Ukraine, 4, Chervonopraporna St., Kharkiv, 61002, Ukraine.}
\author{Y Y Zanimonskiy}
\address{International Slavonic University, 9-A, Otakara Jarosha St., Kharkiv, 61045, Ukraine.}
\begin{abstract}
Relativistic treatment of the finite speed of light correction in absolute gravimeters, as evolved by \emph{Rothleitner and Francis} in \emph{Metrologia} 2011, \textbf{48} 442-445, following the initial publication in \emph{Metrologia} 2011, \textbf{48} 187-195, leads to spurious conclusions. The double Doppler shift implemented in the gravimeters obliterates the difference between its relativistic and non-relativistic formulation. Optical heterodyning used in Michelson-type interferometers makes the quadratic Lorenz-like term of the double Doppler shift discernable against the linear term, while in relativistic experiments the quadratic term has to be detected against the unit. The disturbance of the registered trajectory caused by the finite speed of light includes tracking signal delay as intrinsic part not reducible to the Doppler shifts.
\end{abstract}
%
%
%
%
\newpage
\pagestyle{empty}
\section{Introduction}
Special Theory of Relativity predicts \cite{pauli1981} that measurements performed by instruments at a resting frame would show the increase of mass, contraction of length, and dilation of time as compared to the similar  measurements performed at the frame moving with the velocity $V$. The relative magnitude of these effects is mostly determined by the quadratic term of the Lorenz factor
\begin{equation}
\label{Lorendz}
\gamma = (1-V^2/c^2)^{-\frac12} \approx 1+V^2/2c^2,
\end{equation}
Evanescence of these effects in laboratory settings, along with their unclear nature made interpretation of relativity the subject of active discussions, recently spread to \emph{Metrologia}. The paper \cite{rothleitner2011} claims that the correction for the finite speed of light in absolute gravimeters is the result of relativistic effects. In the reply \cite{rothleitner2011a} to our comment \cite{nagornyi2011d}, where we disagreed with this claim, the authors have brought even more relativistic arguments into the reasoning of the correction. Following the reply \cite{rothleitner2011a}, we address two interrelated questions with far-reaching consequences for physics, geophysics, and metrology:
\begin{itemize}
\item What correction for the finite speed of light should be used?
\item Are relativistic effects observable in modern absolute gravimeters?
\end{itemize}
\section{The gist of the argument: 2 or 3?}
The influence of any phenomenon on the absolute gravimeter measurements can be analyzed in two steps \cite{nagornyi1995}. First, the disturbance of the test mass' tracked trajectory caused by the phenomenon is to be found, then the disturbance has to be translated into the appropriate correction. The comment \cite{nagornyi2011d} discusses problems we found in \cite{rothleitner2011} with both steps of the process, while the reply \cite{rothleitner2011a} addresses only our critique of the first step. The result that the authors are still standing by is the disturbance of the trajectory due to the finite speed of light. The issue actually boils down to the value of $k$ in the expression of the disturbance
\begin{equation}
\label{dg_c_with_k}
\Delta g(t) = \mp k\frac{g_0}{c}(V_0 + g_0t).
\end{equation}
While our analysis \cite{nagornyi2011} reveals that the value of $k$ should be 3, the authors of \cite{rothleitner2011} insist that $k$ equals 2. To investigate the issue, we can simplify (\ref{dg_c_with_k}) by neglecting test mass' initial velocity $V_0$, and its position with respect to the interferometer that determines the sign of (\ref{dg_c_with_k}).

\begin{figure}[h]
\centering
\small
\input{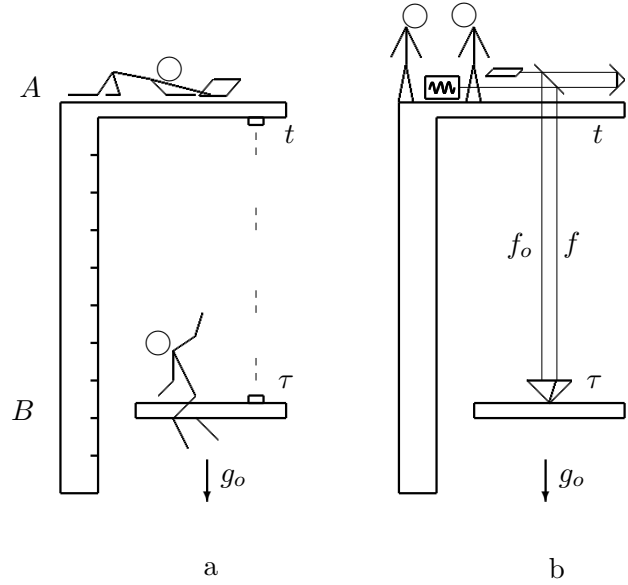}
  \caption[short title]
  {
  \quad\parbox[t]{11cm} {Different models of signal delay in motion tracking:
      \begin{compactdesc}
      \item[a]{--- delay in transferring the information about the coordinate;}
      \item[b]{--- interferometric tracking, equivalent to the delay in the information about velocity.}
      \end{compactdesc}
  }
  }
\label{C-FOLLOWUP}
\end{figure}

Let's follow two physics majors, Alice and Bob, on their quest to measure gravity acceleration. By arrangement with a local amusement park, the students decided to use a ride that has free falling platform. They have marked up the ride with units of length (fig.\ref{C-FOLLOWUP}a)
\label{sec_extra_delay}
and started the experiment. Bob (B) was taking rides, and every time the platform reached next mark, he signaled the event with a flashlight back to Alice (A). As every signal corresponded to a known position of the platform, the students obtained the acceleration by fitting the following model to the data:
\begin{equation}
\label{z_fit}
z = g_0 t^2/2,
\end{equation}
where $z$ was the position of the platform  at the moment $t$, as determined by arrival of the flashlight signals. The students then realized that the moment $t$, at which the signal arrived to the top of the ride was not exactly the same as the moment the signal was send. The delay of the signal traveling from Bob to Alice with the speed $c$ is $|AB|/c$, so the students modified the model as
\begin{equation}
\label{z_fit_corr}
z = g_0 \left(t - \frac{|AB|}{c} \right)^2/2.
\end{equation}
Because $|AB|$ was about the same distance $z$ as in (\ref{z_fit}), they have substituted (\ref{z_fit}) for $z$ in (\ref{z_fit_corr}), and expressed the acceleration like:
\begin{equation}
\label{g_disturb_3}
g(t) = \frac{\D^2 z}{\D t^2}=\frac{\D^2}{\D t^2}\; \left[ g_0 \left(t - \frac{g_0 t^2/2}{c} \right)^2/2 \right]
= g_0 - 3 \frac{g_0^2 t}{c}
.
\end{equation}

Then the students conducted another experiment.
They attached a reflector to the platform and used it as moving arm of the Michelson-type interferometer (fig.\ref{C-FOLLOWUP}b.) Every time the platform advanced one-half of the laser wavelength, the whole period of the interference was observed. But what about the influence of the finite speed of light in this experiment? At the moment the platform moves with the velocity $V$, the frequency of the reflected beam is determined by the double Doppler shift \cite{pauli1981}:
\begin{equation}
\label{f_reflected}
f = f_0 \frac{1-V/c}{1+V/c} \approx f_0 (1-2V/c+2V^2/c^2).
\end{equation}
The observed beat frequency is
\begin{equation}
\label{f_beat}
\Delta f = f_0 - f = 2 f_0 (V/c - V^2/c^2).
\end{equation}
As every period of the beat signal corresponds to the advancement of the test mass on $\lambda/2$, the velocity $\overline V$ deduced from the signal (\ref{f_beat}) is
\begin{equation}
\label{V_beat}
\overline V = \frac{\lambda}{2} \Delta f = \lambda f_0 (V/c - V^2/c^2) = V - V^2/c.
\end{equation}
Substituting $V=g_0t$, the acceleration can be found as
\begin{equation}
\label{g_beat}
g(t) = \frac{\D \overline V}{\D t} = \frac{\D}{\D t} \left[ g_0t - (g_0t)^2/c \right] =
g_0 - 2\frac{g_0^2t}{c}.
\end{equation}
Comparison of (\ref{g_beat}) and (\ref{z_fit_corr}) has puzzled the students, because in the second experiment the finite speed of light yielded the disturbance one-third less then that in the first experiment (\ref{z_fit_corr}). After pondering for a while, the students suspected the source of the problem was with the formula (\ref{f_beat}). ``Look,'' said Alice, ``the difference signal (\ref{f_beat}) emerges at the beam splitter, while the Doppler shifts (\ref{f_reflected}) happen at the reflector.'' ``There is a small, but finite time interval between the moments that same photons touch those two objects,'' added Alice. Following the clue, Bob has modified the formula (\ref{g_beat}) like
\begin{equation}
\label{g_beat_modified}
g(t) = \frac{\D}{\D t}\left[g_0\left(t - \frac{g_0 t^2/2}{c} \right) - \left(g_0\left(t - \frac{g_0 t^2/2}{c} \right)\right)^2/c\right] =
g_0 - 3\frac{g_0^2t}{c}.
\end{equation}
Though the disturbance term became now the same as in formula (\ref{z_fit_corr}), the students were still in doubt. ``Strange thing,''  said Alice, ``the time delay gave us the entire disturbance term when we applied it in the first experiment. And now the same delay has added only one-third of the disturbance...'' ``Maybe, this has something to do with the fact that the delay was applied differently in the second experiment?'' suggested Bob. ``Right!'' exclaimed Alice. ``Look, the first time the delay was applied to the signal carrying the information about the coordinate, and the second time the delay was applied to the velocity signal.'' If in the first experiment Bob could have send the information about the velocity, not coordinate, the model would become
\begin{equation}
\label{V_fit}
V = g_0 t.
\end{equation}
Substituting the delay into this model yields the following acceleration
\begin{equation}
\label{V_fit_corr}
g(t) = \frac{\D V}{\D t}=\frac{\D}{\D t}\;\left[ g_0 \left(t - \frac{g_0 t^2/2}{c} \right)/2\right]
= g_0 - \frac{g_0^2 t}{c}.
\end{equation}
``Amazing!'' said Bob, ``this disturbance term is indeed exactly one-third of that we obtained in the first experiment.'' The students then wrote up a brief summary of what they have learned about the influence of the finite speed of tracking signal on the measured acceleration. Here is the note:

\emph{``Acceleration of the test mass can be obtained from the information of its coordinate or velocity. In either case the tracking signal delivers the information with the same delay. The acceleration disturbance obtained by applying the delay to the velocity signal is one-third of that obtained by applying the delay to the coordinate. The remaining two-thirds of the disturbance come from the double Doppler shift, so that overall disturbance is the same in both cases.''}

The above conclusion states that the more processing of the information is done before its transfer, the smaller the error introduced by the transfer delay. Had the acceleration been obtained right at the test mass, there would be no delay-associated error at all. The acceleration inferred from the velocity (\ref{V_fit_corr}) is three times less prone to the error than that inferred from the coordinate (\ref{g_disturb_3}). As there is still processing involved to convert the velocity into the acceleration, the delay error is unavoidable.

\section{Transverse Doppler shift is impossible in absolute gravimeters}
\begin{figure}[h]
\centering
\small
\input{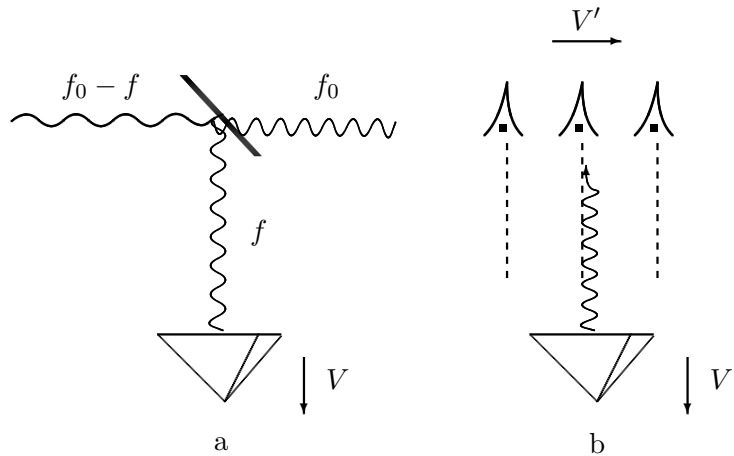}
  \caption[short title]
  {
  \quad\parbox[t]{11cm} {On the possibility of the transverse Doppler shift in absolute gravimeter:
      \begin{compactdesc}
      \item[a]{--- beams producing the interference;}
      \item[b]{--- relative motion of source and observer producing the transverse Doppler shift.}
      \end{compactdesc}
  }
  }
\label{TRANSVERSE}
\end{figure}

The reply \cite{rothleitner2011a} rejects the necessity of the additional delay, leaving the disturbance term like in (\ref{g_beat}). The reasoning against the delay is based on its comparison with the transverse Doppler shift. According to \cite{rothleitner2011a}, the shift takes place in the interferometer, because the beams producing the beat signal meet at the right angle, and one of the beams come from the moving source (fig.~\ref{TRANSVERSE}a). The shift decreases the observed frequency of the signal like
\begin{equation}
\label{f_transverse}
f' = f_0 \sqrt{1-V'^2/c^2} \approx f_0 (1-V'^2/2c^2).
\end{equation}
The authors of \cite{rothleitner2011a} have noticed that the frequency change (\ref{f_transverse}), if doubled and treated like in (\ref{V_beat}) --- (\ref{g_beat}), yields the same additional disturbance (\ref{V_fit_corr}) as the delay. Unfortunately, this intriguing coincidence can not entail any meaningful interpretation, because there are several reasons why the transverse Doppler shift can not exist in absolute gravimeters.
\label{sec_extra_delay}
\begin{enumerate}
\item \emph{Absence of the transverse motion.} The transverse Doppler shift \cite{pauli1981} decreases the observed frequency of a signal issued from the source that moves transversely to the observer's line of sight (fig.~\ref{TRANSVERSE}b). In the interferometer the ``observer'' is the point on the beam splitter where the interference happens. As the test mass moves longitudinally, not transversely with respect to the beam splitter, no transverse shift is possible.
\item \emph{Volatility of the shift.} The observer moving at the right angle to the beam can see the light only during the infinitely small interval of time on which the beam coincides with the observer's line of sight (fig.~\ref{TRANSVERSE}b). Continuous observation of light from moving source is possible only in circular motion. As there is no rotation of the beam splitter in absolute gravimeters, the gedankenexperiment of \cite{rothleitner2011a} in which the transverse Doppler shift is continuously observed during the entire free-fall, is self-contradictory.
\item \emph{Inconsistent disturbances.} The decrease of frequency (\ref{f_transverse}) due to the transverse Doppler shift does not depend on the direction of motion of the light source. Replacing the delay (\ref{V_fit_corr}) with the transverse Doppler shift yields the same disturbance only for free-fall gravimeters with the test mass approaching the beam splitter. For other types of gravimeters, the replacement yields inconsistent results (see Table \ref{tbl_components}). For the upper position of the interferometer, the disturbance magnitude becomes three times less than for the lower position. The symmetric motion does not cancel the disturbances on the upward and downward branches of the trajectory.
\end{enumerate}
Because the transverse Doppler shift can not exist in absolute gravimeters, all and any conclusions based on assuming the opposite, have no ground.
\Table{\label{tbl_components} Components of the trajectory disturbance caused by the finite speed of light, obtained with different treatment of the velocity signal delay:\\
\textbf{Case A}. Traditional treatment \cite{kuroda1991, nagornyi2011}. The sign of the delay follows the sign of the double Doppler shift, yielding opposite disturbances for the upper and lower positions of the interferometer in the free-fall schemas, and zero disturbance for the symmetric schemas.\\
\textbf{Case B}. Treatment via the transverse Doppler shift \cite{rothleitner2011a}. The shift has the same sign for either direction, yielding the disturbance amplitude for the lower position of the interferometer three times greater than that for the upper position in the free-fall schemas. For the symmetric schemas, the disturbances on the upward and downward branches of the trajectory do not cancel each other.\\
(Note. The symbols $\top$ and $\bot$ stand for the upper and the lower positions of the interferometer correspondingly.)
}
\br
& &\centre{6}{Disturbance components}&  \\
 &  & \crule{6} & \\
& &\centre{3}{Approaching beam splitter}&\centre{3}{Receding from beam splitter}&\\
\ns
\ns
Case & Schema & \crule{3} & \crule{3} & Total\\
 & & Double & Velocity  & Transverse  & Double  & Velocity  & Transverse & \\
 & & Doppler & signal  & Doppler  & Doppler  & signal  & Doppler &\\
 & & shift & delay  & shift  & shift   & delay  & shift &\\
\mr
 & free-fall $\top$ & \m\m--- & \m\m--- & \m\m--- & $-2g_0^2t/c$ & $-g_0^2t/c$ & \m\m--- & $-3g_0^2t/c$ \\
A & free-fall $\bot$ & \m$2g_0^2t/c$ & \m$g_0^2t/c$ & \m\m--- & \m\m--- & \m\m--- & \m\m--- & \m$3g_0^2t/c$ \\
 & symmetric $\top$ & \m$2g_0^2t/c$ & \m$g_0^2t/c$ & \m\m--- & $-2g_0^2t/c$ & $-g_0^2t/c$ & \m\m--- & \m\m--- \\
 & symmetric $\bot$ & $-2g_0^2t/c$ & $-g_0^2t/c$ & \m\m--- & \m$2g_0^2t/c$ & \m$g_0^2t/c$ & \m\m--- & \m\m--- \\
\mr
 & free-fall $\top$ & \m\m--- & \m\m--- & \m\m--- & $-2g_0^2t/c$ & \m\m--- & \m$g_0^2t/c$  & $-g_0^2t/c$ \\
B & free-fall $\bot$ & \m$2g_0^2t/c$ & \m\m--- & \m$g_0^2t/c$ & \m\m--- & \m\m--- & \m\m--- & \m$3g_0^2t/c$ \\
 & symmetric $\top$ & \m$2g_0^2t/c$ & \m\m--- & \m$g_0^2t/c$ & $-2g_0^2t/c$ &  \m\m--- & \m$g_0^2t/c$ & \m$2g_0^2t/c$ \\
 & symmetric $\bot$ & $-2g_0^2t/c$ & \m\m--- & \m$g_0^2t/c$ & \m$2g_0^2t/c$ &\m\m --- & \m$g_0^2t/c$ & \m$2g_0^2t/c$ \\
\br
\endTable

\section{Can absolute gravimeters sense relativistic effects?}
Even though the quadratic term of (\ref{f_reflected}) can not be interpreted as transverse Doppler shift, the similarity poses an interesting question. On the one hand, the finite speed of light correction arising from the quadratic term significantly exceeds the sensitivity level of modern instruments \cite{nagornyi2011d}. On the other hand, the change of frequency (\ref{f_transverse}) caused by the transverse Doppler shift is comparable to that causing the correction (\ref{f_reflected}). Does this mean that absolute gravimeters are sensitive enough to detect relativistic effects? At the test mass velocity of 2 ms$^{-1}$, the quadratic term $V^2/2c^2$ of the beat frequency (\ref{f_reflected}) relates to its linear term $V/c$ as $10^{-8}$, which falls within the sensitivity range of modern instruments. In relativistic experiments \cite{chou2010}, the quadratic term has to be compared not to the linear one, but to the unit, which requires much greater sensitivity of $10^{-17}$. The optical heterodyning that produces the beat signal (\ref{f_reflected}) can not be used in relativistic experiments, because the double Doppler shift loses its relativistic properties and becomes the same as non-relativistic one\footnote{Regular double Doppler shift: $(c-V)/c\;\times\;c/(c+V)\equiv (c-V)/(c+V)$,\\
relativistic double Doppler shift: $[((c-V)/(c+V))^{\frac12}]^2 \equiv (c-V)/(c+V)$.}. That's why the corrections used in absolute gravimeters can not be considered as indicative of relativistic effects.

We must admit that sometimes the term ``relativistic correction'' is used pretty liberally, just because the Doppler shift for light (even if it's double shift), or Lorenz algebra is used to analyze the correction \cite{heilmann2000}. We believe that in metrology more conservative approach to naming corrections should be used. Presenting the correction for the finite speed of light in relativistic terms would create unsupported expectations of more physical content in the correction than it actually has.
\section{Conclusions.}
\begin{itemize}
\item Interferometric measurements of acceleration include the finite speed of light disturbance, of which two-thirds are caused by Doppler shifts, and one-third is caused by the velocity signal delay. The disturbance obtained with the delay is consistent with other models of trajectory tracking. The reasoning against the delay based on the transverse Doppler shift does not hold true, as the shift can not exist in absolute gravimeters. The factor $k$ of the disturbance (\ref{dg_c_with_k}) should be 3, the corresponding corrections for different types of instruments are given in \cite{nagornyi2011}.
\item The quadratic term of the double Doppler shift (\ref{f_reflected}) responsible for the finite speed of light correction, is comparable in magnitude to the similar terms of the Lorenz factor (\ref{Lorendz}) and the transverse Doppler shift (\ref{f_transverse}). The similarity does not signify relativistic origins of the correction, because the double Doppler shift is the same in relativistic and non-relativistic formulations. In absolute gravimeters, the quadratic term is compared to the linear term, while in relativistic experiments the quadratic term is compared to the unit. This explains why absolute gravimeters sense the quadratic term, while their sensitivity is much below the level necessary for relativistic experiments.
\end{itemize}
\subsection*{Acknowledgments}
We are very thankful to the authors of \cite{rothleitner2011} for the interesting and though-provoking paper. Analysis of the ideas and results published in the paper enabled much deeper understanding of the finite speed of light correction in absolute gravimeters. Our disagreement with the authors' conclusions by no means depreciates their valuable contribution to the subject.
\section*{References}
%
%

\begin{thebibliography}{10}

\bibitem{pauli1981}
W.~Pauli, {\em Theory of Relativity}.
\newblock New York: Dover, 1981.

\bibitem{rothleitner2011}
C.~Rothleitner and O.~Francis, ``Second-order doppler-shift corrections in
  free-fall absolute gravimeters,'' {\em Metrologia}, vol.~48, no.~3, p.~187,
  2011.

\bibitem{rothleitner2011a}
C.~Rothleitner and O.~Francis, ``Reply to 'comment on second-order
  doppler-shift corrections in free-fall absolute gravimeters','' {\em
  Metrologia}, vol.~48, no.~5, p.~442, 2011.

\bibitem{nagornyi2011d}
V.~D. Nagornyi, Y.~M. Zanimonskiy, and Y.~Y. Zanimonskiy, ``Relativity, doppler
  shifts and retarded times in deriving the correction for the finite speed of
  light: a comment on 'second-order doppler-shift corrections in free-fall
  absolute gravimeters','' {\em Metrologia}, vol.~48, no.~5, p.~437, 2011.

\bibitem{nagornyi1995}
V.~D. Nagornyi, ``A new approach to absolute gravimeter analysis,'' {\em
  Metrologia}, vol.~32, no.~3, pp.~201--208, 1995.

\bibitem{nagornyi2011}
V.~D. Nagornyi, Y.~M. Zanimonskiy, and Y.~Y. Zanimonskiy, ``Correction due to
  the finite speed of light in absolute gravimeters,'' {\em Metrologia},
  vol.~48, no.~3, p.~101, 2011.

\bibitem{kuroda1991}
K.~Kuroda and N.~Mio, ``Correction to interferometric measurements of absolute
  gravity arising from the finite speed of light,'' {\em Metrologia}, vol.~28,
  no.~2, pp.~75--78, 1991.


\bibitem{chou2010}
C.~W. Chou, D.~B. Hume, T.~Rosenband, and D.~J. Wineland, ``Optical clocks and
  relativity,'' {\em Science}, vol.~329, no.~5999, pp.~1630--1633, 2010.

\bibitem{heilmann2000}
R.~K. Heilmann, P.~T. Konkola, C.~G. Chen, and M.~L. Schattenburg,
  ``Relativistic corrections in displacement measuring interferometry,'' {\em
  J. Vac. Sci. Technol. B}, vol.~18, no.~6, pp.~3277--3281, 2000.

\end{thebibliography}

\end{document}